\documentstyle [12pt,twoside]{article}
\begin{document}
\centerline{\Large\bf Violent Relaxation, Phase Mixing, and}
\vskip .1in
\centerline{\Large\bf Gravitational Landau Damping}
\vskip .2in
\centerline{\large\bf Henry E. Kandrup}
\small
\vskip .15in
\centerline{Department of Astronomy and Department of Physics and}
\vskip .05in
\centerline{Institute for Fundamental Theory, University of Florida}
\vskip .05in
\centerline{Gainesville, FL 32611 USA}
\vskip .2in
\begin{abstract}
\par\noindent
This paper outlines a geometric interpretation of flows generated by the 
collisionless Boltzmann equation, focusing in particular on the coarse-grained 
approach towards a time-independent equilibrium. The starting
point is the recognition that the collisionless Boltzmann equation is a 
noncanonical Hamiltonian system with the distribution function $f$ as the 
fundamental dynamical variable, the mean field energy ${\cal H}[f]$ playing
the role of the Hamiltonian and the natural arena of physics being ${\Gamma}$,
the infinite-dimensional phase space of distribution functions. 
Every time-independent equilibrium $f_{0}$ is an energy extremal with respect
to all perturbations ${\delta f}$ that preserve the constraints (Casimirs) 
associated with Liouville's Theorem. If the extremal is a local energy 
minimum, $f_{0}$ must be linearly stable but, if it corresponds instead to 
a saddle point, $f_{0}$ may be unstable. If an initial $f(t=0)$ is 
sufficiently close to some linearly stable lower energy $f_{0}$, its 
evolution can be visualised as involving linear phase space oscillations 
about $f_{0}$ 
which, in many cases, would be expected to exhibit linear Landau damping. If 
instead $f(0)$ is far from any stable extremal, the flow will be more 
complicated but, in general, one might anticipate that the evolution can be 
visualised as involving nonlinear oscillations about some lower energy $f_{0}$.
In this picture, the coarse-grained approach towards equilibrium usually
termed violent relaxation is interpreted as nonlinear Landau damping.
Evolution of a generic initial $f(0)$ involves a coherent initial excitation
${\delta f}(0){\;}{\equiv}{\;}f(0)-f_{0}$, not necessarily small, being 
converted into incoherent motion associated with nonlinear oscillations about 
some $f_{0}$ which, in general, will exhibit destructive interference.
This picture allows for distinctions between regular and chaotic ``orbits''
in ${\Gamma}$: Stable extremals $f_{0}$ all have vanishing Lyapunov exponents,
even though ``orbits'' oscillating about $f_{0}$ may well correspond to 
chaotic trajectories with one or more positive Lyapunov exponents.
\end{abstract}
\vfill\eject
\centerline{\large\bf 1. Introduction and Motivation}
\vskip .1in
\par\noindent
The problem addressed in this paper is how to visualise flows generated by the 
collisionless Boltzmann equation ({\it CBE}), i.e., the gravitational analogue 
of the electrostatic Vlasov equation from plasma physics.

It is generally accepted that many physical problems arising in galactic
dynamics and cosmology can be modeled in terms of the {\it CBE}, perhaps 
allowing also for low amplitude discreteness effects, modeled as friction 
and noise through the formulation of a Fokker-Planck equation, or for a 
coupling to a dissipative fluid described, e.g., by the Navier-Stokes 
equation. Astronomers recognise that an evolution described completely by 
the {\it CBE} is special because of the constraints associated with Liouville's
Theorem, and that, at some level, the flow must be Hamiltonian, which 
precludes the possibility of any pointwise approach towards a time-independent
equilibrium: in the absence of dissipation, one can only speak meaningfully
of a coarse-grained approach towards equilibrium. 
However, there does not seem to be a clear sense of exactly how one ought
to visualise a flow governed by the {\it CBE} or of what sort of 
coarse-graining one ought to implement in order to identify an approach towards
equilibrium.

The conventional wisdom of galactic dynamics (cf. Binney and Tremaine 1987),
as articulated, e.g., by
Maoz (1991), draws sharp distinctions between different aspects of the 
evolution, speaking separately of phase mixing, (linear) Landau damping, and 
violent relaxation. However, such distinctions, even if useful in addressing 
specific physical effects, are arguably {\it ad hoc} and, as such, may obscure 
the overall character of the flow. Plasma physicists (cf. van Kampen 1955, Case
1959) are well acquainted with the fact that, appropriately interpreted, 
linear Landau damping {\it is} a phase mixing associated with the evolution
of a wave packet constructed from a continuous set of normal modes. Moreover,
even though conventional wisdom makes a sharp distinction between violent
relaxation and phase mixing/Landau damping, one can argue that, as is implicit 
in Lynden-Bell's (1967) original paper on violent relaxation, it too 
is a phase mixing process.

The objective here is to present a coherent mathematical description of an
evolution 
described by the {\it CBE} that manifests explicitly the Hamiltonian character 
of the flow. This entails a synthesis and extension of existing work in both
plasma physics and galactic dynamics (cf. Morrison 1980, Morrison and Eliezur
1986, Kandrup 1989, 1998 and numerous references cited therein) which, in the 
context of galactic dynamics, has proven 
useful in understanding problems related to both linear and global stability, 
as well as stability in the presence of weak dissipation (cf. Kandrup 1991a,b, 
Perez and Aly 1996, Perez, Alimi, Aly, and Scholl  1996). Section 2 describes
the precise sense in which the {\it CBE} is an infinite-dimensional 
Hamiltonian system, identifying the natural phase space, exhibiting the 
noncanonical Hamiltonian structure, and then speculating on the possible 
meaning of regular versus chaotic flows.

Section 3 turns to the problem of linear stability for time-independent 
equilibria. This is addressed both in the context of the full noncanonical 
Hamiltonian dynamics and in terms of a simpler canonical Hamiltonian structure 
associated with the tangent dynamics, i.e., identifying explicitly a set of 
canonically conjugate variables in terms of which to analyse linear 
perturbations. One immediate by-product of this discussion is a simple 
explanation (cf. Habib, Kandrup, and Yip 1986) of linear Landau damping which 
manifests explicitly that it is in fact a phase mixing process: Even though a 
perturbation cannot ``die away'' in any pointwise sense, one may expect a 
coarse-grained approach towards equilibrium in which observables like
the density perturbation ${\delta}{\rho}$ eventually decay to zero.

Section 4 generalises the preceding to the case of nonlinear 
stability, allowing for perturbations ${\delta f}$ away from some equilibrium 
$f_{0}$ which are not necessarily small. The intuition derived from that 
problem is then used to motivate one possible way in which to visualise the
flow associated with a generic initial $f(t=0)$. The obvious point is that a 
generic initial $f(0)$ can be viewed as a (possibly strongly nonlinear) 
perturbation of {\it some} equilibrium $f_{0}$, the 
form of which, however, need not be known explicitly. To the extent that this
interpretation is accepted, those aspects of the flow typically denoted 
violent relaxation should be viewed as nonlinear Landau damping/phase mixing 
(cf. Kandrup 1998). Section 5 concludes by describing the mathematical issues 
which must be resolved to make the preceding discussion rigorous and complete. 

A simple mechanical model, which can help in visualising the basic ideas 
described in this paper, is the following: Consider a point particle moving 
in some complicated, many-dimensional potential $V({\bf r})$ which is 
characterised generically by multiple extremal points but which, being 
bounded from below, will have a (in general nondegenerate) global minimum. If
one chooses initial data corresponding to a configuration space point 
${\bf r}$ close to but slightly above some local minimum ${\bf r}_{0}$ and a 
velocity ${\bf v}$ whose magnitude is very small, the subsequent evolution
will involve linear oscillations about ${\bf r}_{0}$, whether or not that point
corresponds to a global minimum. The trajectory of the point particle thus
corresponds to a regular orbit in what appears locally as a harmonic potential.
If the initial deviation from the extremal point becomes somewhat larger, 
because $|{\bf r}-{\bf r}_{0}|$ and/or $|{\bf v}|$ is bigger, one would still 
anticipate oscillations around ${\bf r}_{0}$, but these will now become 
nonlinear and the particle trajectory may well correspond to a chaotic orbit.
Suppose, however, that ${\bf r}_{0}$ is {\it not} the global minimum. In 
this case, one would expect that, for initial data sufficiently far from 
${\bf r}_{0}$, the particle will have left the ``basin of attraction'' 
associated with the local minimum and will instead (generically) exhibit 
strongly nonlinear oscillations about the global minimum (it could of course
oscillate around a different nonglobal minimum!). In the absence of
dissipation, there is no pointwise sense in which the particle evolves towards
the global minimum. However, the nonlinear oscillations in different directions
will in general interfere destructively, so that any initial coherence
between motions in different directions will eventually be lost (at least for
times short compared with the Poincar\'e recurrence time). It is this
loss of coherence which, for the {\it CBE}, gives rise to (linear or nonlinear)
Landau damping.
\vskip .2in
\eject
\centerline{\large\bf 2. The Noncanonical Hamiltonian Formulation}
\vskip .1in
If one considers the Liouville equation appropriate for a collection of 
noninteracting particles evolving in a fixed potential 
${\Phi}({\bf x})$, the natural phase space is the six-dimensional 
phase space associated with the canonical pair $({\bf x},{\bf v})$. If, 
however, one considers the full {\it CBE}, allowing for a self-consistent 
potential ${\Phi}[f({\bf x},{\bf v})]$ determined by the 
free-streaming particles, this is no longer so. In this case, the fundamental
dynamical variable is the distribution function itself, and the natural phase
space ${\Gamma}$ is the infinite-dimensional phase space of distribution 
functions. In general, it is not easy to identify conjugate coordinates and 
momenta 
in this phase space so as to rewrite the {\it CBE} in the form of Hamilton's 
equations. However, one can still capture the Hamiltonian character at a 
formal algebraic level through the identification of an appropriate 
cosymplectic structure (cf. Arnold 1989).\footnote{
One example of a noncanonical Hamiltonian system, well known to astronomers,
is rigid body rotations described by the standard Euler equations
(cf. Landau and Lifshitz 1960). Specifically, as described and generalised, 
e.g., in Kandrup (1990) and Kandrup and Morrison (1993), the Euler equations 
constitute a Hamiltonian system, formulated in the three-dimensional phase
space coordinatised by 
the three components of angular momentum $J_{i}$, ($i=1,2,3$), with the 
Hamiltonian $H[J_{i}]=\sum_{i=1}^{3}J_{i}^{2}/2I_{i}$ (the analogue of eq.
4) defined in terms of the principal moments of inertia $I_{i}$, and the 
Lie bracket (the analogue of eq. 5) given as the natural bracket associated 
with the three-dimensional rotation group, i.e.,
$$[a,b]=\sum_{i,j,k}\,{\epsilon}_{ijk}J_{k}
{\bigl(}{{\partial}a\over {\partial}J_{i}}{\bigr)}
{\bigl(}{{\partial}b\over {\partial}J_{j}}{\bigr)}$$ 
for functions $a(J_{i})$ and $b(J_{i})$. As for the {\it CBE}, there is
also a Casimir (the analogue of eq. 9), namely 
$C[J_{i}]=\sum_{i=1}^{3}J_{i}^{2}$, which restricts motion to the
two-dimensional constant $C$ surface in the three-dimensional phase space.
\par
Astronomers are also acquainted with infinite-dimensional Hamiltonian
systems, at least those realisable in canonical coordinates, one simple
example being the scalar wave equation 
${\partial}^{2}_{t}{\Psi}-{\nabla}^{2}{\Psi}=0$, 
which derives from the Hamiltonian
$${\cal H}={1\over 2}\,
\int\,d^{3}x{\Bigl(}{\Pi}^{2}({\bf x})+|{\nabla}{\Psi}({\bf x}|^{2}{\Bigr)}, $$
where ${\Psi}$ and ${\Pi}$ are canonically conjugate.}

In this context, manifesting the Hamiltonian character of the flow
entails identifying a Lie bracket $[ \, . \, , \, . \, ]$, defined on pairs
of phase space functionals ${\cal A}[f]$ and ${\cal B}[f]$, and a Hamiltonian
functional ${\cal H}[f]$, so chosen that the {\it CBE}
$${{\partial}f\over {\partial}t}+v{\cdot}{{\partial}f\over {\partial}{\bf x}}
-{\nabla}{\Phi}{\cdot}{{\partial}f\over {\partial}{\bf v}}= 0, \eqno(1)$$
with ${\Phi}({\bf x},t)$ the self-consistent potential satisfying
$${\nabla}^{2}{\Phi}=4{\pi}G{\rho}{\;}{\equiv}{\;}\int\,d^{3}v\,f ,\eqno(2)$$
can be written in the form
$${{\partial}f\over {\partial}t}+ {\bigl [}{\cal H},f {\bigr ]} = 0. \eqno(3)$$

The Hamiltonian ${\cal H}$ may be taken as 
$${\cal H}[f]={1\over 2}\,\int\,d{\Gamma}\,v^{2}\,f({\bf x},{\bf v}) -
{G\over 2}\,\int\,d{\Gamma}\,\int\,d{\Gamma}'\,
{f({\bf x},{\bf v})f({\bf x}',{\bf v}')\over |{\bf x}-{\bf x}'|}, \eqno(4)$$
with $d{\Gamma}{\;}{\equiv}{\;}d^{3}xd^{3}v$, 
which corresponds to the obvious mean field energy, as identified, e.g., by 
Lynden-Bell and Sanitt (1969). The bracket is then chosen to satisfy
(Morrison 1980) 
$$[{\cal A},{\cal B}]=\int\,d{\Gamma}\,f\,{\Bigl\{}
{{\delta}{\cal A}\over {\delta}f},{{\delta}{\cal B}\over {\delta}f}{\Bigr\}},
\eqno(5)$$
where $\{g,h \}$ denotes the ordinary Poisson bracket acting on functions 
$g({\bf x},{\bf v})$ and $h({\bf x},{\bf v})$, and ${\delta}/{\delta}f$ 
denotes a functional derivative.
It is straightforward to show that the operation defined by eq. (5) is a skew 
symmetric, bilinear form, satisfying the Jacobi identity
$$[g,[h,k]] + [h,[k,g]] + [k,[g,h]] = 0, \eqno(6)$$
so that it defines a bona fide Lie bracket. However for this bracket one 
verifies immediately that eq. (3) reduces to the {\it CBE} in the form
$${{\partial}f\over {\partial}t}-\{E,f\} = 0,  \eqno(7)$$
where $E$ represents the energy of a unit mass test particle, i.e., 
$$E={1\over 2}v^{2}+{\Phi}({\bf x},t). \eqno(8)$$

A flow governed by the {\it CBE} is strongly constrained by Liouville's
Theorem, which implies the existence of an infinite number of conserved
quantities, the so-called Casimirs ${\cal C}[f]$. Specificially, the flow has 
the property that, for any function ${\chi}(f)$, the value of the phase space 
integral
$$C[f]=\int\,d{\Gamma}\,{\chi}(f) \eqno(9)$$
is invariant under time translation, i.e., $dC/dt=0$. The simplest case 
corresponds to the choice ${\chi}=f$, which leads to conservation of number
(or mass):
$${d\over dt}\,\int\;d{\Gamma}\,f{\;}{\equiv}{\;}0. \eqno(10)$$
By analogy with finite-dimensional systems, where Noether's Theorem relates
conserved quantities to continuous symmetries, these Casimirs reflect internal 
symmetries in the infinite-dimensional phase space ${\Gamma}$ (Morrison and 
Eliezur 1986).

The Casimirs play an important role in analysing the stability of equilibrium
solutions $f_{0}$, where one must restrict attention to perturbations 
${\delta}f$ that satisfy ${\delta}C{\;}{\equiv}{\;}0$ for all possible choices 
of ${\chi}$. As first noted by Bartholomew (1971), this demand implies that 
any allowed perturbation ${\delta}f$ is related to $f_0$ by a canonical 
transformation induced by some generating function $g$, i.e.,
$$f{\;}{\equiv}{\;}f_{0}+{\delta}f={\rm exp}(\{g,\, .\,\})f_{0}. \eqno(11)$$

In addition to the Casimirs, there is also at least one other conserved 
quantity,
namely the mean field energy ${\cal H}[f]$. Specifically, it follows from the 
{\it CBE} that $d{\cal H}/dt{\;}{\equiv}{\;}0$. If one considers initial data 
$f(0)$ characterised by a high degree of symmetry, other conserved quantities 
may also exist. For example, if the initial data correspond to a potential
${\Phi}$ which is spherically symmetric, it follows that the numerical value
of the angular momentum
$${\bf J}{\;}{\equiv}{\;}
\int\,d^{3}xd^{3}v\,f\,{\bf x}{\times}{\bf v} \eqno(12) $$
is necessarily conserved. However, these conserved quantities, if they exist, 
are on a different footing from the Casimirs since they reflect symmetries in 
the particle phase space, rather than internal symmetries associated with the 
infinite-dimensional phase space of distribution functions.

Because of the infinite number of constraints associated with the Casimirs,
the evolution of $f$ is reduced to a lower (but presumably still infinite-)
dimensional phase space hypersurface, say ${\gamma}$. One might 
naively believe that, in the same sense as, e.g., for the Kortweg-de Vries
equation (cf. Arnold 1989), the flow associated with the {\it CBE} is
integrable. In point of fact, however, this is almost certainly not so
(cf. Morrison 1987), the important point being that the Casimirs associated 
with the {\it CBE} are all ``ultralocal'' quantities which do not involve
derivatives of $f$.

At the present time, there is no universally accepted notion of what precisely
one should mean by chaos in an infinite-dimensional Hamiltonian system.
However, one obvious tact entails comparing initially nearby flows and asking 
whether, for some given $f(t=0)$, there exist perturbations ${\delta}f(t=0)$ 
which grow exponentially. This leads naturally\footnote{
I thank Bruce Miller and Klaus Dietz for suggesting this point to me.}
to the notion of a functional 
Lyapunov exponent which, at least formally, can be defined by analogy with the 
definition of an ordinary Lyapunov exponent in a finite-dimensional system
(cf. Lichtenberg and Lieberman 1992). Specifically, given the introduction of 
an appropriate norm $|| \; ||$, one can write 
$${\chi}=\lim_{t\to\infty}\lim_{{\delta}f(0)\to 0} {1 \over t}
\;{||{\delta}f(t)|| \over ||{\delta}f(0) || }. \eqno(13) $$
For finite dimensional systems one knows that, independent of the choice of
norm, the analogue of eq.~(13) will, for a generic phase space perturbation
${\delta}z$, converge towards the largest Lyapunov exponent. Much less is known
about the infinite-dimensional case. For specificity, it thus seems reasonable
to choose $|| \; ||$ as corresponding to a (possibly weighted) $L^{2}$ norm 
defined in the phase space of distribution functions, i.e.,
$$||{\delta}f|| {\;}{\equiv}{\;} \int \, d{\Gamma} \,M({\bf x},{\bf v})\,
|{\delta}f({\bf x},{\bf v})|^{2} , \eqno(14) $$
where $M$ denotes a specified function of ${\bf x}$ and ${\bf v}$. This is,
e.g., the type of norm that has been used in proving theorems about linear
stability.
\vskip .2in
\vfill\eject 
\centerline{\large\bf 3. Linear Stability and Gravitational Landau Damping}
\vskip .1in
The key fact underlying the interpretation of flows described by the 
{\it CBE} and, especially, the problem of stability, is that every
time-independent equilibrium $f_{0}$ is an energy extremal with respect
to ``symplectic'' perturbations ${\delta}f$ of the form (11) which 
preserve the numerical values of every Casimir. This implies that, if one 
restricts
attention to the reduced phase space ${\gamma}$ obtained by freezing the
value of each Casimir at its equilibrium value $C[f_{0}]$, every equilibrium
$f_{0}$ corresponds to an isolated fixed point: To lowest order, the quantity
${\delta}{\cal H}{\;}{\equiv}{\;}0$ for any symplectic ${\delta}f$.  As 
explained below, if $f_{0}$ is a local energy minimum, so that, to next 
leading order, ${\delta}{\cal H}{\;}{\ge}{\;}0$, $f_{0}$ must be linearly 
stable. Alternatively, if 
$f_{0}$ corresponds to a saddle point, so that ${\cal H}$ increases for some
perturbations but decreases for others, linear stability is no longer
guaranteed, although one cannot necessarily infer that $f_{0}$ must be
linearly unstable.

The proof that, to lowest order, ${\delta}{\cal H}$ vanishes for any 
perturbation of the form (11) and the computation of ${\delta}{\cal H}$ to 
higher order are straightforward if one expands (11) perturbatively 
to infer that 
$${\delta}f=\{g,f_{0}\} + {1\over 2} \{g, \{g,f_{0}\} \} +. \, . \, . {\;}
{\equiv}{\;}{\delta}^{(1)}f + {\delta}^{(2)}f + . \, . \, . \, . \eqno(15) $$
It is easy to see that, for any ${\delta}^{(1)}f$, the first variation
${\delta}^{(1)}{\cal H}$ becomes
$${\delta}^{(1)}{\cal H}=\int \, d{\Gamma}\;{\Bigl(}
{1\over 2}v^{2}-G\int\,d{\Gamma}'\, {f_{0}'\over |{\bf x}-{\bf x}'|} {\Bigr)}
{\delta}^{(1)}f = \int\,d{\Gamma}E_{0}{\delta}^{(1)}f , \eqno(16) $$
where $E_{0}$ is the particle energy associated with $f_{0}$.
However, by combining eqs. (15) and (16) and then integrating by parts, one
finds that 
$${\delta}^{(1)}{\cal H}=\int\,E_{0}\,\{g,f_{0}\} = -\int\,d{\Gamma}\,
g\{E_{0},f_{0}\} {\;}{\equiv}{\;} 0 , \eqno(17) $$
where (cf. eq.~7) the final equality follows from fact that $f_{0}$ is 
time-independent.
Extending this calculation to one higher order shows that the second
variation
$${\delta}^{(2)}{\cal H}=-{1\over 2}\,\int\,d{\Gamma}\,
\{g,f_{0}\}\, \{ g,E_{0} \} - {G\over 2}\,\int\,d{\Gamma} \int\,d{\Gamma}'
{ \{g,f_{0}\} \, \{ g',f_{0}' \} \over |{\bf x}-{\bf x}'|} . \eqno(18) $$

To help visualise what is going on, and to understand why linear stability
follows if ${\delta}^{(2)}{\cal H}$ is positive for all symplectic 
perturbations of the form (11), suppose that, in ordinary three-dimensional 
space, the $x$-$y$
plane corresponds to a hypersurface in the reduced ${\gamma}$-space of 
distribution functions.
One can then ``warp'' this plane into a curved two-dimensional surface by
assigning to each $x$-$y$ pair a coordinate $z$ which corresponds to the
numerical value assumed by the energy ${\cal H}$. On this warped surface,
the equilibrium points correspond to those pairs $(x_{0},y_{0})$ which are
extremal in $z$, so that any infinitesimally displaced point 
$(x_{0}+{\delta}x,y_{0}+{\delta}y)$ assumes a new value $z+{\delta}z$. 

If the equilibrium point is a local energy
minimum, any infinitesimal displacement on the surface necessarily increases
the value of $z$, so that, in the neighbourhood of $(x_{0},y_{0})$, the 
surface has the geometry of an upward opening paraboloid. Any perturbation
comes with positive energy and corresponds to bounded motion on the paraboloid.
Thus the equilibrium is linearly stable. In principle, the same conclusion 
also obtains if the extremal point is a local maximum, although one can show 
that,
for realistic equilibria, ${\delta}^{(2)}{\cal H}$ is never strictly negative. 
If, however, the equilibrium corresponds to a saddle point, so that $z$
increases in some directions but decreases in others, the situation becomes
more complicated. In this case, the linearised dynamics implies that it is 
possible to combine a very large negative energy perturbation in one 
direction with a very large positive energy perturbation in another to 
generate a total perturbation with vanishing
energy. In itself, this does not guarantee a linear instability, but the simple
geometric argument for stability that holds for a local minimum is no longer
applicable.\footnote{
Strictly speaking, the application of this finite-dimensional argument to
an infinite-dimensional Hamiltonian system requires that the reduced phase
space ${\gamma}$  be endowed with a metric, so that one knows what is meant 
by distance between points. In practice, this can be done by introducing an
appropriate
$L^2$ norm, which provides the natural extension of the Euclidean notion of
distance to an infinite-dimensional space. In this context, a proof of 
stability entails showing that $||{\delta}f(t)||$ remains bounded for all
times.} 

That saddle points need not 
imply linear instability may seem surprising at first glance. However, the
following two-dimensional example makes clear exactly what can go wrong:
$$H={1\over 2}{\Bigl(}v_{1}^{2}+{\omega}_{1}^{2}x_{1}^{2}{\Bigr)}-
{1\over 2}{\Bigl(}v_{2}^{2}+{\omega}_{2}^{2}x_{2}^{2}{\Bigr)}. \eqno(19) $$
Here $x_{1}=v_{1}=x_{2}=v_{2}=0$ is a time-independent extremal point in the
phase space which corresponds to a saddle but, nevertheless, the equilibrium
is clearly stable. 
This model may seem somewhat contrived but, as discussed in Section V of 
Kandrup and Morrison (1993), such stable saddle points are not uncommon
in various infinite-dimensional Hamiltonian systems.

The preceding argument for stability or lack thereof may seem somewhat unusual 
because it is formulated abstractly in phase space, without the introduction 
of conjugate coordinates and momenta. One might therefore hope that, by 
identifying
an appropriate set of conjugate variables, a more intuitive proof could be
derived. In certain cases, this is in fact possible. One knows that, when
formulated in the full ${\Gamma}$-space, the dynamics cannot be decomposed
completely into canonical variables because of the existence of the Casimirs,
which correspond to null vectors of the cosymplectic structure. If, however,
one passes to the reduced ${\gamma}$ space, where the values of all the 
Casimirs are frozen, one might expect that, at least locally, conjugate 
variables do exist. Indeed, for finite-dimensional systems it follows from 
Darboux's Theorem (cf. Arnold 1989) that, if the cosymplectic structure has 
vanishing determinant, i.e., if there are no null eigenvectors, it is always 
possible to find a set of canonically conjugate variables, at least locally 
(see Section V of Kandrup and Morrison [1993] for a detailed discussion of 
this point). 

One setting in which such a canonical formulation is possible is for the
special case of linear perturbations of an equilibrium $f_{0}$ which is
a function only of the one-particle energy $E$, i.e., $f_{0}=f_{0}(E)$, and
for which the partial derivative 
$F_{E}{\;}{\equiv}{\;}{\partial}f/{\partial}E$ is strictly
negative. Physically the latter restriction implies that the system does not
exhibit a population inversion; mathematically it ensures
that division by $F_{E}$ is well defined. The basic idea, due originally to
Antonov (1960), is to split the linearised perturbation ${\delta}f$ into
two pieces, ${\delta}f_{+}$ and ${\delta}f_{-}$, respectively even and odd 
under a velocity inversion ${\bf v}\to -{\bf v}$, and to view the single 
linearised perturbation equation for ${\delta}f$ as a coupled system for
${\delta}f_{\pm}$. 

When linearised about some equilibrium $f_{0}$, the {\it CBE} reduces to
$${\partial}_{t}{\delta}f - \{E,{\delta}f\} - \{ {\Phi}[{\delta}f],f_{0} \} =0,
\eqno(20) $$
where $E$ is the particle energy associated with $f_{0}$ and
${\Phi}[{\delta}f]$ denotes the gravitational potential ``sourced'' (cf. eq. 2)
by the perturbation ${\delta}f$. If one observes that $E$ is
an even function of ${\bf v}$, that the Poisson bracket is odd under velocity
inversion, and that ${\Phi}[{\delta}f_{-}]$ vanishes identically, it is clear
that eq. (20) is equivalent to the coupled system
$${\partial}_{t}{\delta}f_{+} - \{E,{\delta}f_{-} \}  =0 $$
and
$${\partial}_{t}{\delta}f_{-} - \{E,{\delta}f_{+} \} - 
\{ {\Phi}[{\delta}f_{+}],f_{0} \} =0 .\eqno(21) $$
However, if one differentiates the second of these relations with respect
to $t$, and uses the first to eliminate ${\partial}_{t}{\delta}f_{+}$, it
follows that 
$${\partial}_{t}^{2}{\delta}f_{-} = \{ E, \{ E,{\delta}f_{-} \} \} +
\{ {\Phi}[ \{E,{\delta}f_{-}\}] , f_{0} \} {\;}{\equiv}{\;}F_{E}\,
{\cal A}{\delta}f_{-}, \eqno(22) $$
where ${\cal A}$ denotes a linear operator.
One can then show that, given the identification of ${\delta}f_{-}$ and
${\partial}_{t}{\delta}f_{-}$ as conjugate variables, the equation
$$(-F_{E})^{-1}{\partial}_{t}^{2}{\delta}f_{-} = -{\cal A}{\delta}f_{-} 
\eqno(23)$$
can be derived from the Hamiltonian
\vskip .05in
$$\hskip -2.95in {\widehat{\cal H}}={1\over 2}\int\,{d{\Gamma}\over (-F_{E})}
({\partial}_{t}{\delta}f_{-})^{2} +
{1\over 2}\int\,d{\Gamma}\,{\delta}f_{-}{\cal A}{\delta}f_{-}$$
$$={1\over 2}\int\,{d{\Gamma}\over (-F_{E})}
({\partial}_{t}{\delta}f_{-})^{2} +
{1\over 2}\int\,{d{\Gamma}\over (-F_{E})}
\{E,{\delta}f_{-}\}^{2} - {G\over 2}\,\int\,d{\Gamma}\,\int\,d{\Gamma}'
{ \{E,{\delta}f_{-} \} \, \{E',{\delta}f_{-}' \}\over |{\bf x}-{\bf x}'|}.
\eqno(24) $$
The connection between $\widehat{\cal H}$ and the energy 
${\delta}^{(2)}{\cal H}$ associated with a small symplectic perturbation 
is discussed
in Kandrup (1989). In particular, one can show that ${\widehat{\cal H}}>0$
for all ${\delta}f_{-}$ if and only if ${\delta}^{(2)}{\cal H}>0$ for all
symplectic perturbations.

The fact that ${\cal A}$ is a symmetric (i.e., hermitian) operator facilitates 
a proof that the equilibrium $f_{0}(E)$ is linearly stable if and only 
if ${\widehat{\cal H}}$ (and ${\delta}^{(2)}{\cal H}$) is
positive. Specifically, a simple energy argument (cf. Laval, Mercier, and
Pellat 1965) implies that the magnitude of ${\delta}f_{-}$, and hence 
${\delta}f$, is bounded in time if ${\cal A}$ is a positive operator, so
that ${\delta}^{(2)}{\cal H}>0$, whereas the possibility of perturbations with
$\int\,d{\Gamma}{\delta}f_{-}{\cal A}{\delta}f_{-}<0$ implies the
existence of solutions that grow exponentially. This is easy to understand in 
the language of normal modes. Since ${\cal A}$ is symmetric, it is clear that 
all solutions ${\delta}f_{-}{\;}{\propto}{\;}{\rm exp}(st)$ have $s^{2}$ real, 
so that the evolution is either purely oscillatory or purely 
exponential. If ${\cal A}$ is positive, $s^{2}$ must be negative, so that the
modes are purely oscillatory. If, however, ${\cal A}$ is not a positive 
operator, there exist modes with $s^{2}>0$, which implies an exponential
instability.\footnote{
In point of fact, one anticipates that, for this simple case, 
${\cal A}$ is guaranteed to be positive. It is believed (cf. Binney and
Tremaine 1987) that any $f_{0}$ depending only on $E$ corresponds to a 
spherically symmetric configuration; but assuming that the mass density
${\rho}$ associated with $f_{0}$ is spherical one can prove that ${\cal A}$
is indeed positive (cf. Kandrup 1989).}

This sort of normal mode expansion facilitates a simple geometric picture
of an infinite-dimensional configuration space of perturbations 
${\delta}f_{-}$ which is (locally) embeddable in the reduced ${\gamma}$-space.
The equilibrium $f_{0}$, which is necessarily an extremal point of the
full Hamiltonian ${\cal H}$, satisfies 
${\delta}f_{-}{\;}{\equiv}{\;}{\partial}_{t}{\delta}f_{-}{\;}{\equiv}{\;}0$.
An arbitrary initial perturbation entails a kinetic energy ${\cal K}=
\int\,d{\Gamma}(-F_{E})^{-1}({\partial}_{t}{\delta}f_{-})^{2}$ which is
necessarily positive and a potential energy ${\cal W}=
\int\,d{\Gamma}{\delta}f_{-}{\cal A}{\delta}f_{-}$ whose sign depends on
the properties of ${\cal A}$. If ${\cal A}$ is a positive operator, the
evolution in configuration space involves a particle with ``mass'' 
$(-F_{E})^{-1}$ moving in an
infinite-dimensional harmonic potential which corresponds to an upwards 
opening paraboloid. Linear stability is therefore assured. If, however, 
${\cal A}$ is not always positive, ${\delta}f_{-}{\;}{\equiv}{\;}0$ 
corresponds to a
saddle point, rather than a local minimum, and the flow is linearly unstable.

In visualising all of this, there is the strong temptation to think of the
normal modes as being discrete, i.e., corresponding to honest square integrable
eigenfunctions rather than singular eigendistributions. This, however, is
not necessarily justified. 

Assuming completeness, one can always view any linear perturbation of an
equilibrium $f_{0}(E)$ with $F_{E}<0$ as a 
superposition of normal modes, writing ${\delta}f$ as a formal sum
$${\delta}f({\bf x},{\bf v},t)=\sum_{\sigma}\;
A_{\sigma}g_{\sigma}({\bf x},{\bf v}){\rm exp}(i{\sigma}t) , \eqno(25) $$
where $g_{\sigma}$ labels the eigenvector, ${\sigma}$ is the corresponding
frequency, which is necessarily real, and $A_{\sigma}$ is an expansion 
coefficient.\footnote{
Strictly speaking, this sum must be interpreted (cf. Riesz and Nagy 1955) as
a Stiltjes integral.} Modulo largely unimportant technical details, the modes
then divide into two types, namely: (1) a countable set of discrete frequencies
belonging to the point spectrum, for which the corresponding eigenvectors are 
well-behaved (e.g., square-integrable) eigenfunctions; and (2) a continuous
set of frequencies belonging to the continuous spectrum, for which the 
eigenvectors are singular eigendistributions. 

The distinction between these two types of modes is extremely important (cf.
Habib, Kandrup, and Yip 1986). Because true eigenfunctions are
nonsingular, they can in principle be triggered individually, i.e., one can
choose a reasonable initial ${\delta}f$ which populates only a single discrete
mode. By contrast, because eigendistributions are singular, one cannot sample
a single continuous mode. Rather, any smooth ${\delta}f$ sampling the
continuous spectrum must really be constructed as a wavepacket comprised of
a continuous set of modes. The important point then is that, when evolved 
into the future, such a wavepacket implies a damping of coarse-grained 
observables like the density ${\rho}$. In other words, if the modes are
continuous there is a precise sense in which the perturbation ``dies away''
and the system exhibits a coarse-grained approach towards the original
equilibrium $f_{0}$.

The physics here is analogous to what arises in ordinary quantum mechanics.
If, in that setting, one considers a physical observable like angular
momentum with a discrete spectrum, one can construct well behaved
eigenstates which, when evolved into the future, maintain their coherence
for all time: the only effect of the evolution is a coherently oscillating
phase. If, however, one considers an observable like position or linear
momentum, where the spectrum is continuous, this is no longer so. In this 
case, a normalisable initial state must be constructed from a continuous set
of singular eigendistributions, so that the best one can do is build a
localised (e.g., minimum uncertainty) wavepacket. However, when evolved into 
the future such a wavepacket will necessarily spread because different 
eigendistributions have different phase velocities. 

It is this loss of coherence associated with the spreading of a wavepacket
that corresponds to (linear) Landau damping. In the context of plasma physics,
Landau damping was derived originally (Landau 1946) in a very different way, 
through the introduction of a Fourier-Laplace transform and an analysis of 
poles in the complex plane. However, at least for the electrostatic Vlasov 
equation (cf. Case 1959), i.e., the electrostatic analogue of the {\it CBE}, 
the mathematical equivalence of these two pictures of Landau damping is well 
understood. 
The physics underlying their equivalence is discussed in Kandrup (1998).

For the special case of perturbations of an homogeneous neutral plasma 
characterised by an isotropic distribution of velocities that is everywhere 
nonvanishing, the modes can be computed explicitly (cf. Case 1959), and one 
finds generically that 
$$g_{\sigma}({\bf x},{\bf v})={\rm exp}(i{\bf k}{\cdot}{\bf x})\;
g_{\sigma}({\bf v}) , \eqno(26) $$
where $g_{\sigma}({\bf v})$ is a singular eigendistribution involving a
Dirac delta. In this setting, an examination of the perturbation associated
with a given ${\bf k}$-vector at a fixed phase space point 
$({\bf x}_{0},{\bf v}_{0})$ yields no evidence of damping away. 
Rather, one finds persistent oscillations
${\propto}{\;}{\rm exp}[i{\bf k}{\cdot}({\bf x}_{0}-{\bf v}_{0}t)].$
This is simply a manifestation of the fact that, without the introduction
of some coarse-graining, one cannot speak of the system returning to
equilibrium. If, however, a coarse-graining is implemented by integrating over 
any finite range of velocities, one discovers that the resulting
$\int\,d^{3}v\,{\delta}f({\bf x},{\bf v},t)$ will in fact damp away.

The obvious question, therefore, is: will perturbations ${\delta}f$ of
a generic equilibrium solution to the {\it CBE} correspond to discrete modes, 
continuous modes, or a combination of both? Unfortunately, this is a difficult 
question to answer. It appears impossible to calculate the modes explicitly
for realistic equilibria, and a formal analysis is also difficult because the 
operator ${\cal T}$ entering into the linearised equation 
${\partial}_{t}{\delta}f={\cal T}{\delta}f$ is not elliptic and involves
a singular integral kernel. However, the normal modes can, and have, been
computed for a variety of nontrivial equilibrium solutions to the corresponding
electrostatic Vlasov equation (cf. van Kampen 1955, Case 1959), and the 
results derived thereby would seem suggestive.

Perhaps the most important result derived for the Vlasov equation is that
discrete modes are seemingly the exception, rather than the norm, arising
only if the equilibrium in question manifests nontrivial boundary
conditions, e.g., the existence of a maximum speed $v_{m}$ such that
$f({\bf v}){\;}{\equiv}{\;}0$ for $|{\bf v}|>v_{m}$. In particular, one
can prove that the modes are always purely continuous if $f_{0}$ is an
analytic function of ${\bf v}$ in the complex plane.\footnote{
The validity of Landau's original derivation of exponential damping actually
relies on the implicit assumption that $f_{0}$ is analytic. If it is not, his 
manipulations of contours and evaluation of poles cannot be justified.}
The best known example of a nonempty point spectrum is the case of so-called 
van Kampen (1955) modes, which arise precisely in those configurations where 
there is maximum velocity. In the usual interpretation (cf. Stix 1962), Landau
damping is understood as resulting from a resonance between ``particles''
(the unperturbed $f_{0}$) propagating with velocity ${\bf v}$ and a ``wave''
(the perturbation ${\delta}f$) that propagates with phase velocity 
${\bf c}$. Discrete van Kampen modes correspond to perturbations which
propagate with a phase velocity ${\bf c}$ for which $f_{0}({\bf c})=0$, so
that no resonance is possible.

By analogy, one might therefore conjecture (cf. Habib, Kandrup, and Yip 1986)
that, for the gravitational {\it CBE}, most perturbations will in fact
correspond to continuous modes that damp away, but that some perturbations,
especially longer wavelength disturbances that probe the phase space 
boundaries of the system, could in fact correspond to discrete modes. In this 
connection, it is interesting to note that there do in fact exist exact 
time-dependent solutions 
to the {\it CBE}, seemingly appropriate for a system like a galaxy, that 
exhibit finite amplitude undamped oscillations about some time-independent 
$f_{0}$ (cf. Louis and Gerhart 1988, Sridhar 1989). The interesting point, 
then is that in all these models the time-independent $f_{0}$ contains
phase space ``holes,'' i.e., regions in the middle of the occupied phase space 
region where $f_{0}\to 0$. Whether these sorts of solutions are generic, and
whether they could arise from reasonable initial conditions, is at the present
unclear.

Finally, it should be noted that, in point of fact, one can in principle
get (at least temporary) phase mixing or loss of coherence even for the much 
simpler
case of a finite set of discrete modes. For example, if one considers the 
function $x(t)= \sum_{p=10}^{29}\rm{cos}(0.1pt)$ over the finite interval 
$0<t<1000$, one infers a rapid damping of the initial coherent excitation with 
$x=20$ to a much smaller value oscillating about $x=0$ with 
typical amplitude $|x|{\;}{\sim}{\;}1$. If, however, the evolution is tracked
for a somewhat longer time one finds that the initial coherence is regained. 
An infinite
set of continuous modes differs from this toy model in two important ways,
namely (1) the recurrence time is infinitely long and (2) it is impossible 
to consider a smooth initial excitation that does not damp.

\vskip .2in
\centerline{\large\bf 4. Nonlinear Stability and Global Evolution}
\vskip .1in
Suppose, once again, that attention is focused on some linearly stable 
equilibrium $f_{0}(E)$ with $F_{E}<0$, but that one is now interested in the 
effects of larger perturbations ${\delta}f$, i.e., 
the problem of nonlinear stability.
To the extent that the normal modes of the linear problem remain complete,
one can still envision evolution in terms of these modes, the important point,
however, being that, because of nonlinearities, the modes will now interact. 
This is, e.g., the basis for the standard quasilinear analyses implemented
in plasma physics, which allow for the effects of the quadratic term
${\nabla}{\Phi}{\cdot}({\partial}{\delta}f/{\partial}{\bf v})$ which is 
ignored when considering linear perturbations. 

Mode-mode couplings are important in that they facilitate the transfer of 
energy between different modes, which makes the physics more 
complicated. However, one might still anticipate that, if the modes are 
continuous, Landau damping can and will occur. Because of the interactions 
between modes, the simple model of a dispersing quantum mechanical wavepacket 
is no longer directly applicable, but the basic phenomenon of loss of 
coherence is robust. Indeed, there are many examples in nonlinear dynamics
of flows satisfying nonlinear evolution equations where phase mixing occurs. 
It thus seems reasonable to suppose that, when considering the nonlinear 
evolution of some perturbation ${\delta}f$, one will encounter nonlinear 
Landau damping. For the case of an electrostatic plasma, nonlinear Landau 
damping is a well known, and reasonably well understood, phenomenon (cf. 
Davidson 1972 and references cited therein). Indeed, there are simple 
geometries where the nonlinear evolution can be computed explicitly in the 
context of a systematic perturbation expansion, thus facilitating analytic 
formulae for exactly how this phenomenon works (cf. Montgomery 1963).

Mode-mode couplings can also lead to another important possibility, namely
the onset of chaos. Because $f_{0}$ is a local energy minimum, one
knows that any infinitesimal perturbation ${\delta}f$ will simply oscillate,
each eigenvector corresponding to motion in a ``direction'' in configuration
space that is orthogonal to the motion of all the other eigenvectors. This
implies that, for the fixed point $f_{0}$, the Lyapunov exponents, which were 
defined in eq. (13) as probing the average linear instability of the orbit
generated from some initial $f(0)$, must all vanish identically. One might
anticipate further that, when evolved into the future, other phase space 
points 
sufficiently close to $f_{0}$ will also correspond to regular orbits with
vanishing Lyapunov exponents. Thus, e.g., for finite-dimensional systems one
knows that there is a regular phase space region of finite measure surrounding 
every stable periodic orbit. However, for sufficiently large ${\delta}f$,
where mode-mode couplings become significant and the motion cannot be well
approximated by orthogonal harmonic oscillations, one might anticipate that
many, if not all, perturbations will evolve chaotically. If true, this would 
suggest that a ``typical'' perturbation with ${\delta}{\cal H}=
{\cal H}[f_{0}+{\delta}f]-{\cal H}[f_{0}]$ will evolve ergodically on (some
subset of) the constant energy hypersurface in the ${\gamma}$-space with
energy ${\cal H}[f_{0}+{\delta}f]$.

This idea of the onset and development of chaos is an infinite-dimensional 
generalisation
of what is typically found when considering the motion of a point mass in a 
multi-dimensional nonlinear potential which has only one extremal point, 
a global minimum.\footnote{Even order truncations of the Toda
potential (cf. Kandrup and Mahon 1994) provide a simple two-dimensional 
example.} Low energy orbits sufficiently close to the pit of the potential 
move in what is essentially a harmonic potential, so that their motion is 
regular. If, however, the energy is raised one finds generically that, unless 
the motions in different directions remain completely decoupled, there is
an onset of global stochasticity which leads, for sufficiently high energies,
to well developed chaotic regions.

This configuration space description is not appropriate when considering
generic equilibria, where the energy $\widehat{\cal H}$  associated with a 
small perturbation cannot be written easily as a functional of conjugate 
variables, and there is no guarantee that $\widehat{\cal H}$ can be written
as a simple sum of kinetic and potential contributions, ${\cal K}$ and
${\cal W}$. Modulo technical details, one might expect that canonical 
phase space coordinates do exist, at least in principle, but the energy
$\widehat{\cal H}$ associated with the tangent dynamics could in general be
an arbitrary quadratic functional $\widehat{\cal H}[q,p]$ of the conjugate 
variables $q$ and $p$. Moreover, even for
the simple model of an equilibrium $f_{0}(E)$ with $F_{E}<0$, it may not be
possible to extend the canonical description to allow for arbitrarily large
perturbations ${\delta}f$. One really needs to return to a full phase space
description.

As discussed in Section 3, if for some equilibrium $f_{0}$ the second 
variation ${\delta}^{(2)}{\cal H}$ is positive for all ${\delta}f$, a
linearised perturbation corresponds in phase space to stable motion on an
upwards opening infinite-dimensional paraboloid. As long as this surface
remains convex, one would anticipate that stability will persist and, as such,
one would expect intuitively that the equilibrium could remain nonlinearly
stable even for small but finite ${\delta}f$. The normal modes of the 
linearised problem become coupled, but the geometric argument for stability
should remain valid. In particular, one can presumably visualise the evolution 
of ${\delta}f$ as involving nonlinear {\it phase space} oscillations about
the equilibrium point $f_{0}$. 

If, however, $f_{0}$ corresponds to a stable
saddle, one might suppose that even the smallest nonlinearities could trigger
an instability (cf. Moser 1968, Morrison 1987). Thus, e.g., for the simple 
toy model of two stable oscillators
described by eq. (19), it is possible to trigger an instability by introducing
even very tiny mode-mode couplings which allow energy to be transferred
between modes. Indeed, as noted by Cherry (1925), if the two frequencies
are in an appropriate resonance, e.g., ${\omega}_{2}^{2}=2{\omega}_{1}^{2}$,
the introduction of a simple cubic coupling implies that initial data 
arbitrarily close to $x_{1}=v_{1}=x_{2}=v_{2}=0$ can lead to solutions in
which $x_{1}$, $x_{2}$, $v_{1}$, and $v_{2}$ all diverge in a finite time.
If true, this expectation about saddle points would suggest that, even though 
they can
be linearly stable, they cannot represent reasonable candidate equilibria
in terms of which to model real astronomical objects.

If a linearly stable $f_{0}$ corresponds to a unique extremal point in the
${\gamma}$-space, the surface which near $f_{0}$ is a paraboloid will 
remain upwards opening even if ${\delta}f$ is very large, so that stability
should persist for arbitrarily large perturbations. In other words,
one would expect that the equilibrium $f_{0}$ is globally stable: In this
case, any phase space deformation ${\delta}f$ increases the energy, and the 
evolution of an initial ${\delta}f(0)$ will involve nonlinear phase space 
oscillations around the unique stable fixed point.

If, however, there exist multiple extremal points in the ${\gamma}$-space, 
each corresponding to a local energy minimum, the situation is much more
complicated. In this case, one would anticipate that, for sufficiently large
${\delta}f$, the distribution function can actually be transferred from the
``basin of attraction'' of one equilibrium $f_{0}$ to the ``basin'' of some
other $f_{1}$. In other words, the evolution of ${\delta}f(0)$ could yield
oscillations
around $f_{1}$, rather than $f_{0}$. By suitably fine-tuning the perturbation,
one can in principle displace the system from any one basin to any other. 
However, by analogy with the behaviour observed in finite-dimensional systems,
one might expect generically that, if the perturbation is sufficiently large,
its motion can be interpreted as involving nonlinear phase space 
oscillations about the global
energy minimum. To the extent that this is true, one would anticipate that a
sufficiently large perturbation will tend generically to push $f$ into the
``basin of attraction'' of the equilibrium $f_{0}$ that corresponds to a global
energy minimum.

If one considers an initial perturbation ${\delta}f(0)$ that is sufficiently 
large, the subsequent evolution will in general be almost completely unrelated 
to the initial equilibrium $f_{0}$ and, as such, the way in which one
visualises the evolution ${\delta}f(0)$ is really no different from the way
in which one can, and arguably should, envision the evolution of a generic 
$f(0)$. In
other words, the physical picture described above can be used equally well
to visualise generic flows associated with the initial value problem, the 
only difference being that, in general, one may know nothing at all about 
what time-independent equilibria $f_{0}$ actually exist.

Specification of an initial $f(0)$ fixes the values of all the Casimirs for
all times, thus determining ${\gamma}$, the reduced infinite-dimensional 
phase space which constitutes the natural arena of physics. This $f(0)$ also
fixes the numerical value of the conserved energy ${\cal H}$ and, as
such, determines the constant energy hypersurface in the ${\gamma}$-space
to which the flow is necessarily restricted. By analogy with finite-dimensional
Hamiltonian systems (cf. Kandrup and Mahon 1994) one might expect that, when
evolved into the future, $f(0)$ will exhibit a coarse-grained approach
towards an invariant measure on this hypersurface, i.e., a suitably defined
microcanonical distribution. If the flow associated with $f(0)$ is chaotic,
one might anticipate an approach towards this invariant measure that is
exponential in time. If, alternatively, the flow is regular, one might instead
expect a power law approach. However, in either case one might anticipate an
approach towards a ``phase-mixed'' invariant measure. In this context, the
crucial question is then: to what extent can this invariant measure be 
interpreted as corresponding to a distribution function $f$ executing phase
space oscillations about one or more equilibrium solutions $f_{0}$? 

It is easy to see that, in the ${\gamma}$-space, there must exist one or more 
extremal 
points with ${\delta}^{(1)}{\cal H}=0$, these corresponding to equilibrium
solutions $f_{0}$ for which all the Casimirs share the same values as the
Casimirs associated with $f(0)$. Indeed, one knows that, for sufficiently
smooth initial data, the {\it CBE} admits global existence (cf. Pfaffelmoser 
1992, Schaeffer 1991), so that ${\delta}f$ cannot diverge and, presumably,
the Hamiltonian is bounded from below. However, this implies that there must 
exist at least one $f_{0}$, namely the global energy minimum (although in 
principle the global minimum could be degenerate). The question therefore
becomes: in the basin of which $f_{0}$ (or $f_{0}$'s) does the flow reside?

In principle, the evolved distribution function $f$ could execute phase space
oscillations about any $f_{0}$ with lower energy, which one presumably
depending on the initial $f(0)$. However, one might conjecture
that, if the initial $f(0)$ is sufficiently far from any equilibrium
$f_{0}$, it will execute oscillations around the global minimum $f_{0}$.
The initial $f(0)$ cannot exhibit a pointwise approach towards this, or any
other, $f_{0}$. However, one might expect that, in general, the initial
deviation ${\delta}f(0)=f(0)-f_{0}$ will exhibit nonlinear Landau damping
so that, in terms of observables like the density ${\rho}$, ${\delta}f$
does indeed ``die away,'' and one can speak of a coarse-grained approach
towards the equilibrium $f_{0}$.
\vskip .2in
\centerline{\large\bf 5. Conclusions and Unanswered Questions}
\vskip .1in
The aim of this paper is to suggest a potentially fruitful way in which to
visualise flows described by the {\it CBE} and, in particular, the
expected coarse-grained approach towards an equilibrium. No claim is made
regarding mathematical rigor, and it is not clear that all the details are 
completely correct. However, the viewpoint developed here does have the
advantage that it incorporates what {\it is} known rigorously about the
{\it CBE}, and that it provides a framework in terms of which to pose precise,
well defined questions.
In this context, there are at least three basic questions which, if answered
satisfactorily, would yield important insights into the physical properties
of a flow generated by the {\it CBE}:
\vskip .02in
\par\noindent
1. Will generic initial conditions exhibit effective Landau damping, thus 
allowing one to speak of an efficient coarse-grained evolution towards some
equilibrium $f_{0}$? In the context of linear Landau damping, the answer to
this question depends on the spectral properties of the linearised evolution
equation. If the modes are all continuous, every initial perturbation will
eventually phase mix away, so that physical observables like the density will
damp to zero. If, however, some of the modes are discrete, it is possible to
construct initial perturbations that do not damp away. At the present time,
it is not clear whether, for realistic galactic models, the spectrum is purely
continuous, although the investigation of various toy models is currently
underway (Lynden-Bell 1997, private communication). 

To the extent that $N$-body simulations are reliable and that, for 
sufficiently large $N$, they capture the same physics as the {\it CBE},
the fact that most initial conditions yield an efficient approach towards some
statistical equilibrium can be interpreted as evidence that nonlinear Landau
damping is in general very effective. However, there {\it do} exist toy models
like one-dimensional gravity where one ends up with undamped oscillations. For 
example, the evolution of counterstreaming initial conditions in 
one-dimensional systems (either gravitational or electrostatic) can lead to a 
final state which corresponds seemingly to a distribution function $f$ 
exhibiting finite amplitude undamped oscillations about a (near-) equilibrium 
$f_{0}$ (cf. Mineau, Feix, and Rouet 1990). This toy model actually 
corroborates the physical intuition described in this paper in the sense that, 
as one would expect, the phase space contains a large ``hole,'' i.e., a region 
where $f_{0} \to 0$. Whether or not analogous results obtain for two- and
three-dimensional systems is as yet unclear, although the problem is currently
under investigation (Habib, Kandrup, Pogorelov, and Ryne, work in progress).
\vskip .02in
\par\noindent
2. Are functional Lyapunov exponents the ``right'' way in which to identify
chaos in infinite-dimensional systems and, assuming that they are, will a
generic flow associated with the {\it CBE} be chaotic? Given this definition,
will standard results from finite-dimensional chaos remain at least
approximately valid?
Although not proven for generic finite-dimensional systems, there is the
physical expectation that, when evolved into the future, a chaotic initial
condition will evolve towards an invariant distribution on a time scale 
that is related somehow to the spectrum of Lyapunov exponents. This implies
however that, at asymptotically late times, one can visualise the flow as 
densely filling a chaotic phase space region of finite measure. Assuming,
however, that this is true, the Ergodic Theorem provides important information
about the statistical properties of the flow, implying the equivalence of
time and phase space averages (cf. Lichtenberg and Lieberman 1992).

One other point about chaos in the {\it CBE} should be stressed: The 
definition proposed in 
this paper is, at least superficially, completely decoupled from the (also
interesting) question of whether individual orbits in a self-consistent
potential generated from the {\it CBE} are, or are not, chaotic. This latter
question refers to the behaviour of nearby trajectories in the six-dimensional
particle phase space. The ``natural'' definition of chaos for the {\it CBE}
should presumably reflect properties of the flow in the infinite-dimensional
phase space of distribution functions.
\vskip .02in
\par\noindent
3. For a specified initial $f(0)$, towards which equilibrium $f_{0}$ will
the system evolve? 
Given $f(0)$, one can compute the numerical value of all possible Casimirs,
thus identifying explicitly the ${\gamma}$-space to which the evolution is
restricted. The obvious problem, then, is to identify all time-independent 
equilibria $f_{0}$ in ${\gamma}$ and to determine which initial conditions 
correspond
to which equilibria. Although unquestionably difficult, this is a problem 
that is both well defined mathematically and well motivated physically. 
Finding all equilibria is equivalent mathematically to finding all extremal 
points in ${\gamma}$. However, to the extent that one chooses to visualise 
the flow as involving oscillations in the ${\gamma}$-space, there is no 
question physically but that the extremal points
define ``basins of attraction'' associated with the oscillations.

The basic points described in this paper are easily summarised: 
\par\noindent
1. The {\it CBE} is a Hamiltonian system, albeit an unusual one. The
fundamental dynamical variable is the distribution function $f$, not the
particle ${\bf x}$'s and ${\bf v}$'s; and it is not always possible (at least
easily) to identify canonically conjugate variables. 
\par\noindent
2. Because the {\it CBE} is Hamiltonian, there can be no pointwise approach
towards equilibrim. The best for which one can hope is a coarse-grained
approach towards equilibrium. 
\par\noindent
3. Even though the phase space ${\gamma}$ associated with the dynamics is 
infinite-dimensional, one might expect that much of one's intuition from
finite-dimensional systems remains valid. In particular, one might anticipate
an asymptotic approach towards an invariant measure, and one might hope to make
meaningful distinctions between regular and chaotic flows. 
\par\noindent
4. The phenomenon normally designated as linear Landau damping can be 
interpreted as a phase mixing of a continuous set of normal modes. Whether
a small initial perturbation will always eventually Landau damp/phase mix 
away depends on whether the normal modes for the linearised perturbation 
equation are discrete or continuous.
\par\noindent
5. To the extent that one's ordinary intuition about finite-dimensional
phase spaces remains approximately valid, the evolution of generic initial
data should be interpreted as involving nonlinear (phase space) oscillations
about one or more energy extremals, which correspond to time-independent
equilibria $f_{0}$. The phenomenon of violent relaxation should thus be 
interpreted as nonlinear phase mixing/Landau damping which, if efficient, 
will facilitate a coarse-grained approach towards equilibrium.
\vskip .2in
\centerline{\large\bf Acknowledgments}
\vskip .1in
\par\noindent
I am pleased to acknowledge useful discussions, collaborations, and
correspondence with Salman Habib, Donald Lynden-Bell, Bruce Miller, Phil 
Morrison, and Daniel Pfenniger. I am grateful to Barbara Eckstein for 
comments on the exposition. Work on this manuscript began while I was
a visitor at the Observatoire de Marseille, where I was supported by the
{\it C.N.R.S.}, and continued during a visit to the Aspen Center for Physics.
Limited financial support was provided by the National Science Foundation
grant PHY92-03333.
\vfill\eject
\par\noindent
Antonov, V. A. 1960, Astr. Zh. {\bf 37}, 918 (English trans. in Sov. Astr. -- 
AJ {\bf 4}, 859 [1961]).
\par\noindent
Arnold, V. I. 1989, Mathematical Methods of Classical Mechanics, 2nd ed.,
Springer-Verlag, Berlin.
\par\noindent
Bartholomew, P. 1971, MNRAS {\bf 151}, 333.
\par\noindent
Binney, J. and Tremaine, S. 1987, Galactic Dynamics, Princeton University 
Press, Princeton.
\par\noindent
Case, K. M. 1959, Ann. Phys. (NY) {\bf 7}, 349.
\par\noindent
Cherry, T. M. 1925, Trans. Cambridge Philos. Soc. {\bf 23}, 199.
\par\noindent
Davidson, R. C. 1972, Methods of Nonlinear Plasma Theory, Academic Press,
New York.
\par\noindent
Habib, S., Kandrup, H. E., and Yip, P. F. 1986, Ap. J. {\bf 309}, 176.
\par\noindent
Kandrup, H. E. 1990, Ap. J. {\bf 351}, 104.
\par\noindent
Kandrup, H. E. 1991a, Ap. J. {\bf 370}, 312.
\par\noindent
Kandrup, H. E. 1991b, Ap. J. {\bf 380}, 511.
\par\noindent
Kandrup, H. E. 1998, Ann. N. Y. Acad. Sci., in press.
\par\noindent
Kandrup, H. E. and Mahon, M. E. 1994, Phys. Rev. {\bf E 49}, 3735.
\par\noindent
Kandrup, H. E. and Morrison, P. J. 1993, Ann. Phys. (NY) {\bf 225}, 114.
\par\noindent
Landau, L. D. 1946, Soviet Phys. - JETP {\bf 10}, 25.
\par\noindent
Landau, L. D. and Lifshitz, E. M. 1960, Mechanics, Addison-Wesley, Reading, MA.
\par\noindent
Laval, G., Mercier, C., and Pellat, R. 1965, Nucl. Fusion {\bf 5}, 156.
\par\noindent
Lichtenberg, A. J. and Lieberman, M. A. 1992, Regular and Chaotic Dynamics,
Berlin, Springer, 2nd ed.
\par\noindent
Louis, P. D. and Gerhart, O. 1988, MNRAS {\bf 233}, 337.
\par\noindent
Lynden-Bell, D. 1967, MNRAS {\bf 136}, 101
\par\noindent
Lynden-Bell, D. and Sanitt, N. 1969, MNRAS {\bf 143}, 167.
\par\noindent
Maoz, E. 1991, Ap. J. {\bf 275}, 687.
\par\noindent
Mineau, P., Feix, M. R., and Rouet, J. L. 1990, Astron. Astrophys. {\bf 228}, 
344.
\par\noindent
Montgomery, D. 1963, Phys. Fluids {\bf A6}, 1109.
\par\noindent
Moser, J. K. 1968, Mem. Am. Math. Soc. {\bf 81}, 1.
\par\noindent
Morrison, P. J. 1980, Phys. Lett. {\bf A 80}, 383.
\par\noindent
Morrison, P. J. 1987, Z. Naturforsch. {\bf 42a}, 1115.
\par\noindent
Morrison, P. J. and Eliezur, S. 1986, Phys. Rev. {\bf A 33}, 4205.
\par\noindent
Perez, J. and Aly. J.-J. 1996, MNRAS {\bf 280}, 689.
\par\noindent
Perez, J., Alimi, J.-M., Aly, J.-J., and Scholl, H. 1996, MNRAS {\bf 280}, 700.
\par\noindent
Pfaffelmoser, K. 1992, J. Diff. Eqns. {\bf 95}, 281.
\par\noindent
Riesz, F. and Sz.-Nagy, B. 1955, Functional Analysis, Ungar, New York.
\par\noindent
Schaeffer, J. 1991, Commun. Part. Diff. Eqns. {\bf 16}, 1313
\par\noindent
Sridhar, S. 1989, MNRAS {\bf 201}, 939.
\par\noindent
Stix, T. H. 1962, The Theory of Plasma Waves, McGraw-Hill, New York.
\par\noindent
van Kampen, N. G. 1955, Physica {\bf 21}, 949.
\end{document}